\documentclass[sn-mathphys-num]{sn-jnl}

\usepackage{graphicx}%
\usepackage{multirow}%
\usepackage{amsmath,amssymb,amsfonts}%
\usepackage{amsthm}%
\usepackage{mathrsfs}%
\usepackage[title]{appendix}%
\usepackage{xcolor}%
\usepackage{textcomp}%
\usepackage{manyfoot}%
\usepackage{booktabs}%
\usepackage{algorithm}%
\usepackage{algorithmicx}%
\usepackage{algpseudocode}%
\usepackage{listings}%

\theoremstyle{thmstyleone}%
\theoremstyle{thmstyletwo}%

\theoremstyle{thmstylethree}%

\begin{document}

\title[High-performance in-vacuum optical system for quantum optics experiments in a Penning-trap]{High-performance in-vacuum optical system for quantum optics experiments in a Penning-trap}


\author*[1]{\fnm{Joaqu{\'i}n} \sur{Berrocal}}\email{jberrocal@ugr.es}

\author[1,2]{\fnm{Daniel} \sur{Rodr{\'i}guez}}\email{danielrodriguez@ugr.es}

\affil*[1]{\orgdiv{Departamento de F{\'i}sica At{\'o}mica, Molecular y Nuclear}, \orgname{Universidad de Granada}, \city{Granada}, \postcode{18071}, \country{Spain}}

\affil[2]{\orgdiv{Centro de Investigaci{\'o}n en Tecnolog{\'i}as de la Informaci{\'o}n y las Comunicaciones}, \orgname{Universidad de Granada}, \city{Granada}, \postcode{18071}, \country{Spain}}



\abstract{Accurate measurements with implications in many branches in Physics have been accessed using Penning traps and conventional techniques within a temperature regime where each eigenmotion of a charged particle is still a classical harmonic oscillator. Cooling the particle directly or indirectly with lasers allows reaching the quantum regime of each oscillator, controlling subtle effects in the precision frontier by detecting photons instead of electric current. In this paper, we present a new in-vacuum optical system designed to detecting \mbox{397-nm} fluorescence photons from individual calcium ions and Coulomb crystals in a 7-T Penning trap. Based on the outcome of computer simulations, our design shows diffraction-limited performance. 
The system has been characterized using a single laser-cooled ion as a point-like source, reaching a final resolution of 3.69(3)~$\mu$m and 2.75(3)~$\mu$m for the trap's axial and radial directions, respectively, after correcting aberrations.}




\maketitle

\section{Introduction}\label{sec1}

Quantum optics experiments with cooled trapped ions rely on optical imaging systems for the detection of the fluorescence photons emitted by the ions~\cite{leibfried2003}. Typically, an internal electric-dipole transition of the ion with a lifetime in the order of nanoseconds is addressed by laser light and a fraction of the emitted photons is collected by the optical system and focused on photon detectors such as electron-multiplying charge-coupled devices~(\mbox{EMCCDs}) and photomultiplier tubes~(PMTs)~\cite{neuhauser1978,wineland1978}. A high numerical aperture~(NA), which allows collecting a few percent of the emitted fluorescence, is essential to achieve a fast detection~(in the order of milliseconds) at the single particle level in experiments involving non-destructive internal state discrimination~\cite{nagourney1986}. 

Customized optical systems consisting of multiple lenses to minimize aberrations~\cite{alt2002} are popular in cold-atom experiments and radiofrequency/Paul traps. The typical limited size of these setups allows placing the objective outside vacuum so that a proper alignment can be performed to achieve diffraction-limited performance. Comparatively, Penning traps based on superconducting magnets are subject to spatial constrains due to their large volume and the limited size of the bore.  In the framework of this work, fluorescence detection at the single ion level has only been demonstrated in other Penning-trap experiments devoted to quantum simulation~\cite{gilmore2021}, quantum computation~\cite{hrmo2019,jain2024}, or new schemes to measure fundamental constants~\cite{cornejo2021}. Opposite to these Penning-trap experiments, this optical system has been implemented in vacuum in a beamline-like setup \cite{berrocal2022} which includes external ion sources to create any species of interest.

Our main motivation is to perform Penning-trap motional-frequency spectroscopy in the quantum regime \cite{cerrillo2021}, 
to outperform the detection based on monitoring the electric current a single charged particle induces on the trap electrodes~\cite{wineland1975}. Although this technique and its variants have yielded remarkable results for atomic mass determination in various fields~\cite{cornell1990,gabrielse1999,rainville2004,vandyck2006,borchert2022,medina2023,schweiger2024}, photon detection could still be advantageous due to its universality in terms of mass-to-charge ratio and the ultra-low motional energies required~\cite{rodriguez2012}. This has been recently demonstrated in our group for a single ion cooled to the Doppler limit using the high-resolution optical system described in this paper~\cite{berrocal2024}. The eigenmotions of the ion were individually probed by an external electric field and its motional amplitude was measured directly from the photon distribution determined with the EMCCD or from the photon counting with a PMT. The higher spatial resolution of the optical system leads to reduced systematic uncertainties. 

In this paper, we first present the design guidelines of the new optical system developed for the Penning-trap experiment at the University of Granada. We then discuss the results of the simulations, which indicate diffraction-limited performance. We also present the characterization of the system in terms of resolution using square impulse targets. Finally, we show the analysis of images of a single laser-cooled ion at the Doppler limit~(point-like source) using wave-aberration retrieval techniques to evaluate the performance of the system once it was installed in vacuum.

\section{Design and simulations}

\begin{figure}
\centering
\includegraphics[width=\linewidth]{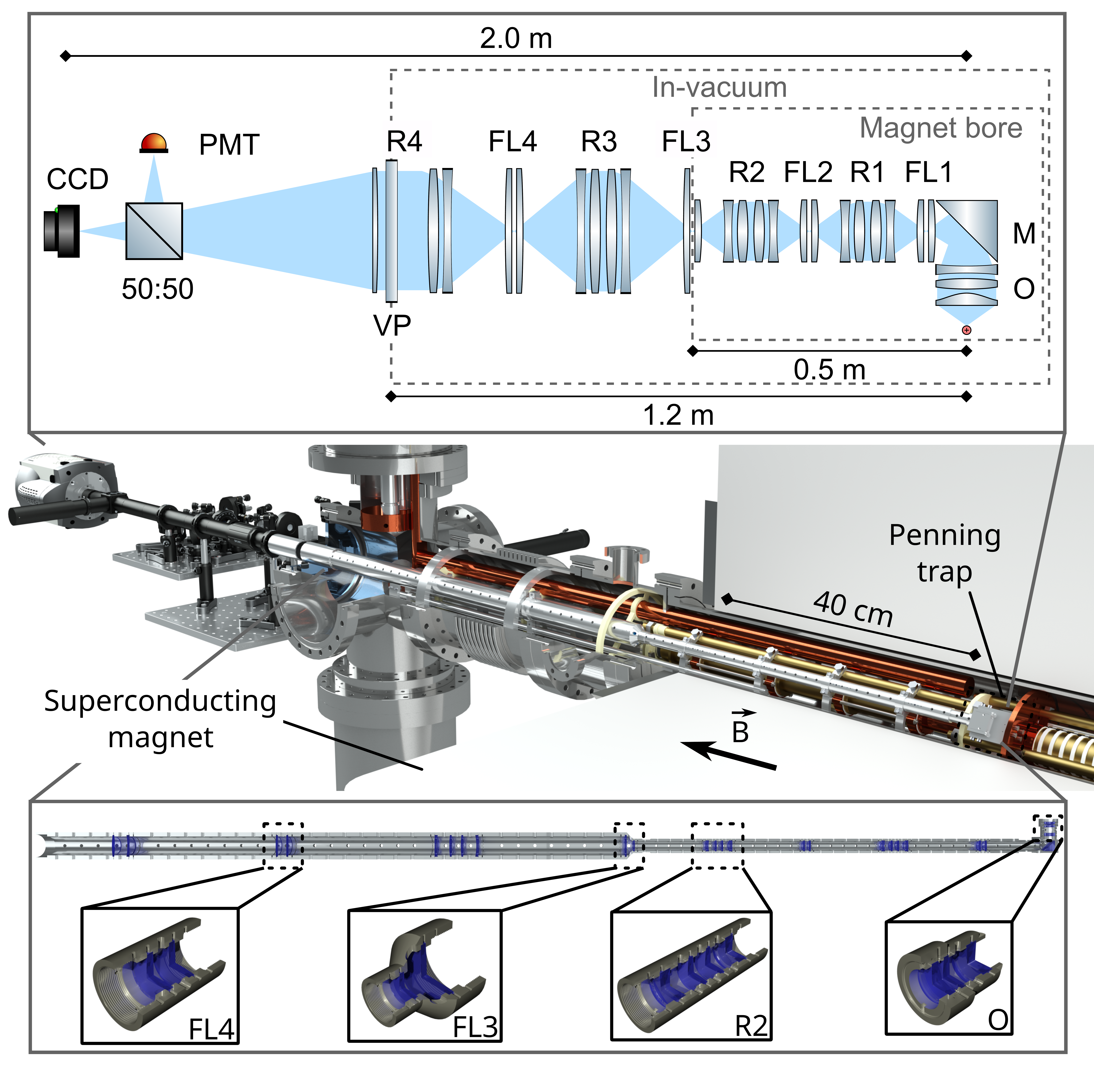}
\caption{The optical system in the Penning-trap beamline. Center: rendered three-dimensional drawing including the 7-T superconducting magnet and the rest of the setup. Top: cross-section sketch (not-to-scale). Bottom: rendered two-dimensional cut where several distinctive elements are zoomed in. O:~objective; M:~mirror; FL:~field lens; R:~relay; VP:~viewport.}
\label{figure_1}
\end{figure}

Figure~\ref{figure_1} shows the conceptual design of the optical system and its arrangement in the experimental setup~(a detailed description of the Penning-trap beamline can be found elsewhere~\cite{gutierrez2019b,berrocal2022}). Most of it~(about 1.2~m) operates under UHV conditions, being rigidly anchored to the Penning-trap tower. It consists of an objective~(O) that provides ${5\times}$ and a series of relays~(R1-R4) that collimate the image without magnification. Field lenses~(FL1-FL4) consisting of pairs of plano-convex lenses centered on the image plane are used to avoid vignetting~\cite{malacara}. A final factor of ${15\times}$ is achieved by adding a single lens outside vacuum. This value is high enough so that two ions are imaged a few pixels apart when projected onto the EMCCD sensor~(Andor iXon Ultra 888, ${13\times13}$~$\mu$m). The relays are doublets of plano-concave and bi-convex commercial lenses in which the relative distance was optimized to achieve diffraction-limited performance. All the lenses are commercial models made from \mbox{N-BK7} or Fused Silica and have anti-reflective coating for ${\lambda=397}$~nm.

The objective is based on an aspheric lens. Other alternatives based on diffraction-limited four-lens objectives~\cite{alt2002,gempel2019,ball2019} were discarded because of the little space available in vacuum. It is comprised of \mbox{1/2''-diameter} lenses, whose specifications are detailed in Appendix~\ref{appendix_tables}. The relative distances between the different elements were optimized using Zemax~(Zemax OpticStudio\textsuperscript{\textregistered} 2020), for which the wave aberration, defined as the optical path difference between the ideal paraxial wavefront and the real one, was used as the figure of merit to minimize. First, the focal point of the asphere was determined for ${\lambda=397}$~nm, resulting in a working distance of 21.963~$\mu$m. Then, the relative distance in the doublet consisting of the bi-convex and the plano-concave lenses was optimized. The gap between the asphere and the doublet was chosen arbitrarily~(5~mm in this case), and the field lens was likewise placed at a certain distance from the image plane. The effect the latter has on the image formation is minimal, and it only affects as a defocus. The optimization routine was run again for the two previously varied distances. Near on-axis diffraction-limited performance was achieved in all the cases, which was observed in the modulation transfer function~(MTF) and wave-aberration plots.

For the relays simulations, a sequential procedure similar to that explained for the objective's was followed~(specifications in Appendix~\ref{appendix_tables}). The final part of the optical system, called R4 in Fig.~\ref{figure_1}, consists of half of a one-inch relay plus a single lens with a focal length of ${f'\approx750}$~mm. Although the latter is a simple plano-convex lens, it does not introduce further significant aberrations due to its small NA.

The optical mounts are custom made and the distance between the elements is maintained using hollow tubes, with very-fine threads to assure a good alignment. 
All the pieces are perforated to pump the inner volume of the tubes. The optimal lens spacings found in the simulations are physically defined by the lens separators, having these the shape of the corresponding spherical lens so that the contact is made on the full contour of the surface for better alignment.

\begin{figure}[t]
\centering
\includegraphics[width=\linewidth]{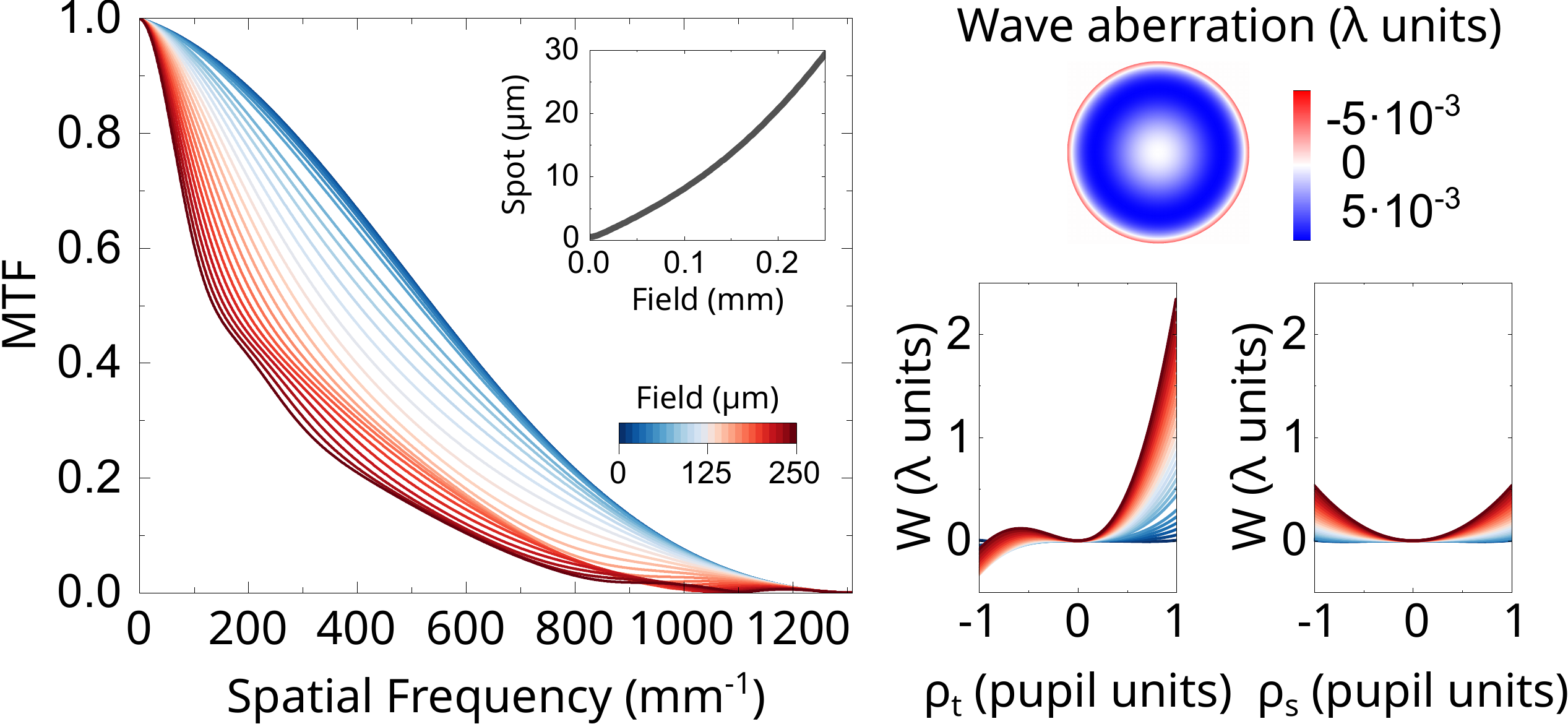}
\caption[Simulation of the optical system]{Simulation of the optical system. Left:~MTF's tangential component for an object placed at different field positions and RMS spot radius~(inset). The axis of abscissa represents the spatial frequency in the object plane. Top right:~colour map representation of the on-axis wave aberration in wavelength units. Bottom right:~tangential~(left) and sagittal~(right) cross sections of the wave aberration for an object placed at different field positions. The field coordinate is given in units of the exit pupil, defined as the image of
the aperture stop formed by the part of the system which follows it.}
\label{figure_2}
\end{figure}

Figure~\ref{figure_2} shows some of the simulation results for the entire optical system. The performance of the optical system is computed based on the wave aberration~(right side of Fig.~\ref{figure_2}). This function can be calculated analytically if one knows the shape of the lenses, listed in Appendix~\ref{appendix_tables}. Although the system is diffraction-limited for on-axis performance, it cannot be moved once installed under UHV, so it is important to know its off-axis performance. For this purpose, the impulse was placed at different field distances from the optical axis on the object plane. The MTF and the root mean square~(RMS) spot radius~(inset) show that a position deviation of 200~$\mu$m, corresponding to a wave aberration of 1.5~$\lambda$, is the maximum affordable deviation to resolve around 10~$\mu$m. In the case of the modulation, a value of 0.2 corresponds to the Rayleigh criterion for two-point resolution~(the first intensity minimum of each point lays on the maximum of the other).

\section{Out-of-vacuum performance analysis}

The response of the system to square-wave impulses was investigated outside vacuum using a~1951~USAF test target~(Thorlabs Inc. R1DS1P). The object, placed in the objective's focal plane, consisted of two sets of three parallel bars, oriented in two perpendicular directions and separated by different distances. The optical system was fixed to an optical table, and the resolution target was mounted on a three-axis stage so that the region of interest could be placed at the focal point. The camera used was a scientific complementary metal-oxide-semiconductor~(sCMOS) image sensor~(Andor Zyla 4.2). The target was illuminated with \mbox{397-nm} light, and measurements were performed on groups 5, 6, and 7. The magnification obtained was ${16.8(4)\times}$. By measuring the maximum and minimum intensities on the sCMOS sensor, $I_{\mathrm{max}}$ and $I_{\mathrm{min}}$, it was possible to determine the contrast transfer function~(CTF) as
\begin{equation}
\mathrm{CTF}=\frac{I_{\mathrm{max}}-I_{\mathrm{min}}}{I_{\mathrm{max}}+I_{\mathrm{min}}}.
\end{equation}
The CTF is the analogous function to the MTF but for square impulses.

If the set of square impulses is chosen as a complete and orthogonal basis, the CTF can be calculated as the inner product of the object and the image, both normalized to the total intensity. Another way of evaluating the performance is to work on the Fourier basis so that the convolution theorem can be applied,
\begin{equation}
\label{eq:opt_sys_deconvolution}
\left|\mathcal{F}\lbrace I_i\rbrace\right| = \mathrm{MTF} \cdot \left|\mathcal{F}\lbrace I_g\rbrace\right|,
\end{equation}
where $I_{g}$ and $I_{i}$ are the intensities in the object and image planes, respectively.

\begin{figure}[t]
\centering
\includegraphics[width=\linewidth]{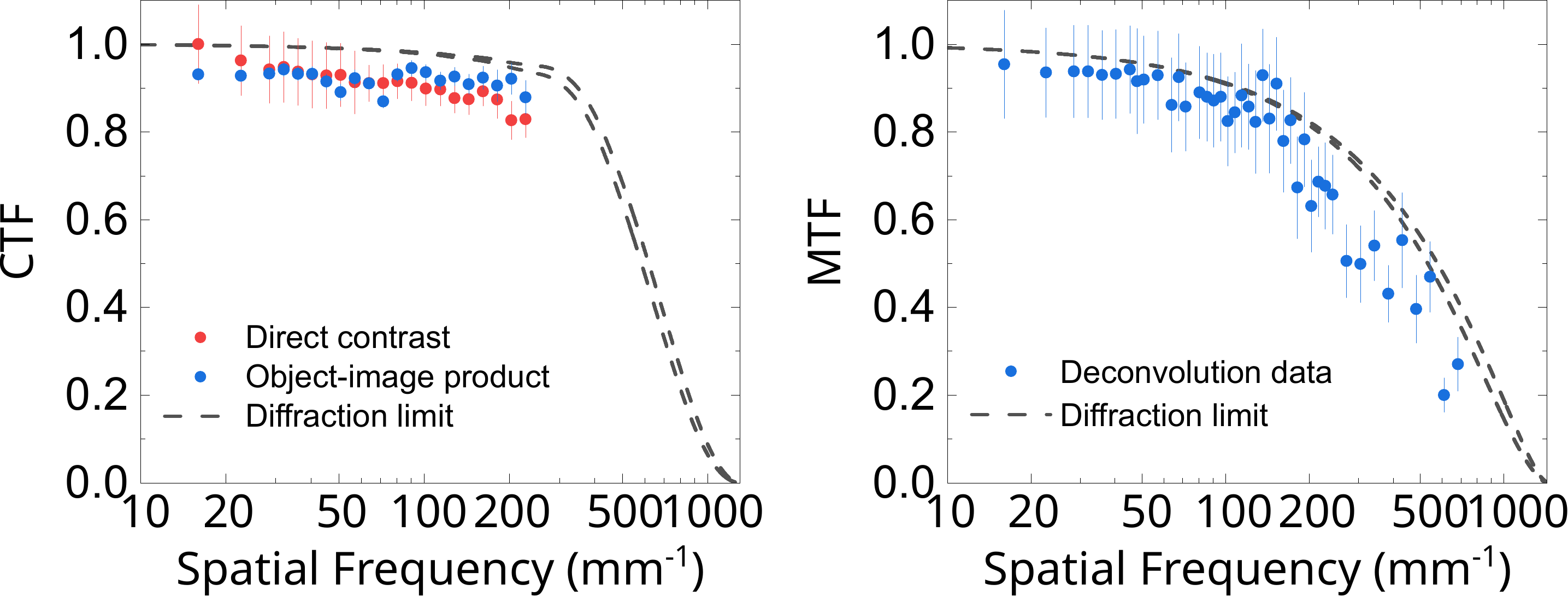}
\caption[Frequency analysis of the measurements based on the 1951~USAF resolution target]{Frequency analysis of the measurements based on the 1951~USAF resolution target. The displayed data are the average of the performance for each two perpendicular bar sets of this test target. Left:~CTF measured directly and from the image-object product. Right:~MTF calculated using the convolution theorem. The lower grey dashed line includes the sensor~(\mbox{6.5~$\mu$m~$\times$~6.5~$\mu$m}) frequency transmission.}
\label{figure_3}
\end{figure}

Figure~\ref{figure_3} shows the performance of the optical system as a function of the spatial frequency of the object. The square-impulse analysis is shown on the left side. Similar results are obtained using the CTF and the MTF, being the performance close to the diffraction limit down to at least~5~$\mu$m of line separation. A deeper insight into higher frequencies was reached by employing the method using the deconvolution theorem to measure the system's~MTF. In the analysis shown on the right side of Fig.~\ref{figure_3}, the fundamental frequency and the third harmonic were used, obtaining reliable results. From the relationship between the modulation and the Rayleigh criterion, the optical system is capable of resolving two points separated by only 1.5~$\mu$m. Using Eq.~(\ref{eq:opt_sys_deconvolution}) for higher harmonics resulted in large uncertainties.

\section{In-vacuum performance analysis}

A single trapped ion cooled to the Doppler limit is a perfect target to test the optical system's point spread function~(PSF). In the case of the Penning trap, the axial frequency was~${\omega_{z}=2\pi\times333}$ kHz, which implies a motional amplitude ${z_{\mathrm{max}}\approx300}$~nm for~$^{40}$Ca$^+$ at 1~mK~(Doppler limit). Since~${z_{\mathrm{max}}<\lambda}$, the ion can thus be considered as a good approximation to a point-like source. If the temperature of the ion was higher, these results would represent the worst-case scenario for the performance of the optical system.

The measured intensity PSF,~${I_{\mathrm{PSF}}\equiv\left|\mathrm{PSF}\right|^{2}}$, can be directly related to the Zernike polynomials~(see Eqs.~(\ref{eq:opt_sys_psf}) and~(\ref{eq:opt_sys_zernike1}) in Appendix \ref{appendix_theory}). Taking into account that Fraunhofer diffraction is a Fourier transform,
\begin{equation}
I_{\mathrm{PSF}} = \left|\mathcal{F}\lbrace e^{-ik\sum_{m,n}c_{nm}Z_{n}^{m}}\rbrace\right|^{2}.
\end{equation}
Since the modulus is not a bijective function, a direct decomposition of~$I_{\mathrm{PSF}}$ into the basis of the Zernike polynomials is not possible. Fitting algorithms have been developed to find the best approach \cite{iglesias1998,barakat1992}, which has already been used in a Paul trap~\cite{wong2016}.

\begin{figure}
\centering
\includegraphics[width=\linewidth]{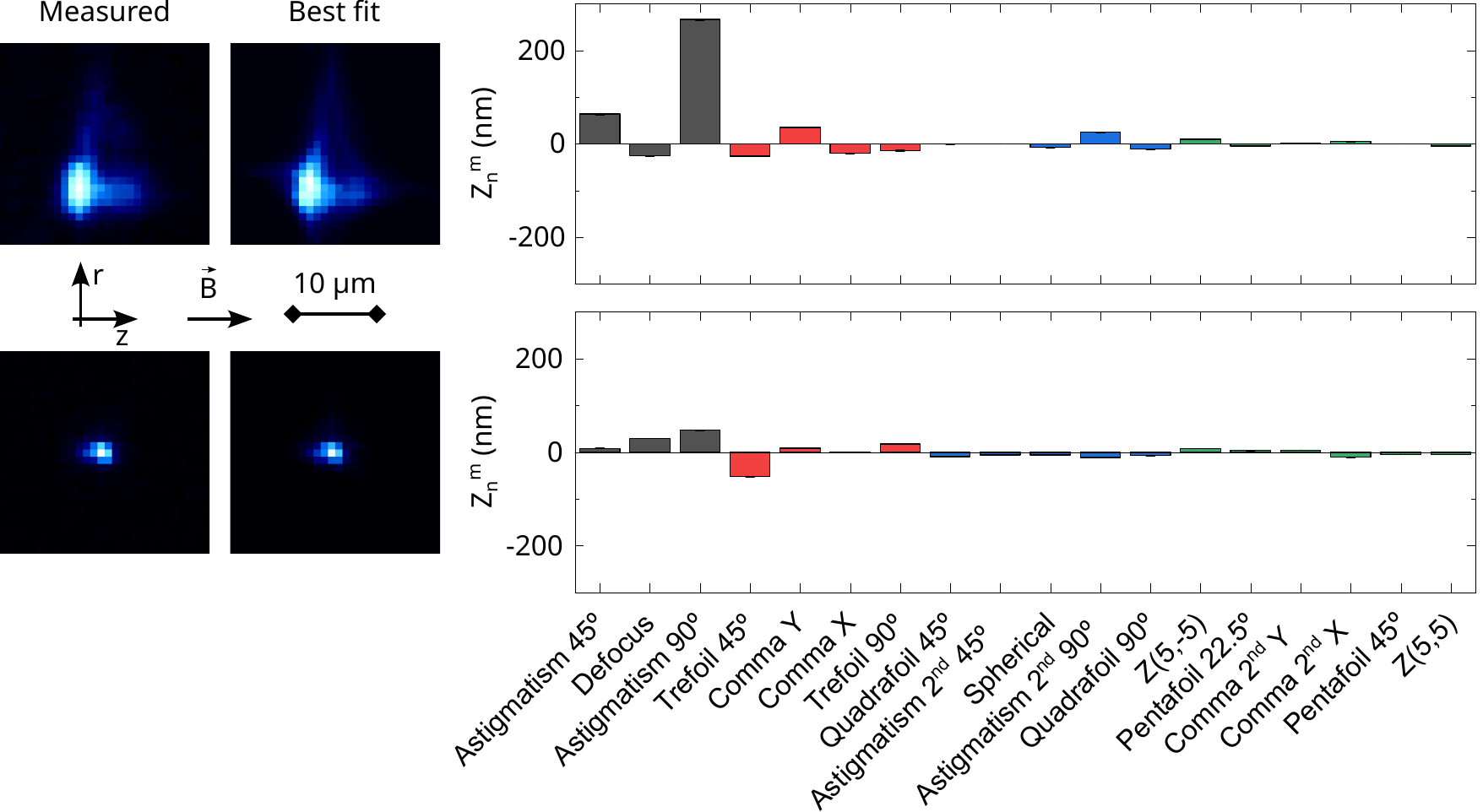}
\caption{Wave aberration analysis. The data shown in the upper/lower panels was taken before/after implementing the external correction consisting of a cylindrical lens. Both the measured image and the best fit are shown on the left side. The accumulation time for the measured image is 1~s. The weight of the first 21~(piston, X-, and \mbox{Y-tilt} are omitted) Zernike polynomials used in the best fit are shown on the right side. The different colours stand for the five orders in the radial coordinate~(see Eq.~(\ref{eq:opt_sys_zernike2}) in Appendix~\ref{appendix_theory}).}
\label{figure_4}
\end{figure}

The outcome from the analysis of the ion image is shown in the upper panels of Fig.~\ref{figure_4}. The magnification was calculated from the ion-ion distance $d_{\mathrm{ion-ion}}$ in a two-ion Coulomb crystal~\cite{james1998},
\begin{equation}
\label{eq:crystal_equilibrium_distance}
d_\mathrm{ion-ion} = \sqrt[3]{\frac{q^2}{2\pi\varepsilon_0m_s\omega_z^2}},
\end{equation}
where $q$ is the ion charge, $\varepsilon_0$ is the vacuum permittivity, $m$ is the ion mass, and $\omega_z$ is the common axial frequency of the balanced two-ion crystal. The value obtained is ${15.4(1)\times}$, in agreement with the design specifications. Compared to the diffraction-limited case, for which ${\mathrm{FWHM}\approx0.8}$~$\mu$m, the measured image was much broader and strongly asymmetric. A total of 21 Zernike polynomials, up to fifth order in the radial coordinate, were used in the fitting procedure, obtaining a reduced \mbox{chi-square} of ${\chi_{\nu}=3.45}$. Introducing sixth-order Zernike polynomials did not improve the results significantly.

Decomposing the image into Zernike polynomials revealed a strong astigmatism, noticeable even in its second-order component. The second most prominent component was comma. Both types of aberration appear when the object is not placed on the optical axis. This is consistent with a misalignment of the optical system with the trap due to machining tolerances at the various anchoring points along the Penning-trap tower. The working distance was correct, as the magnification matches the design value.

Astigmatism can be easily identified by the appearance of two different focal positions for two perpendicular directions due to the different radius of curvature seen by an off-axis object. Following the results shown in the upper part of Fig.~\ref{figure_4}, a focus scan was performed around the circle of least confusion, observing a difference of 25~mm between two focus points for the trap's axial and radial directions. This suggested a lateral shift in the position of the optical system from the trap center. Astigmatism can be corrected with a cylindrical lens, which provides focusing only in one direction. A negative lens with ${f'=-400}$~mm was chosen, so that the first~(tangential) focus point was forced to coincide with the second~(sagittal). It was placed about 110~mm behind the image plane and the final configuration was found experimentally by minimizing the ion's spot radius.

\begin{figure}
\centering
\includegraphics[width=\linewidth]{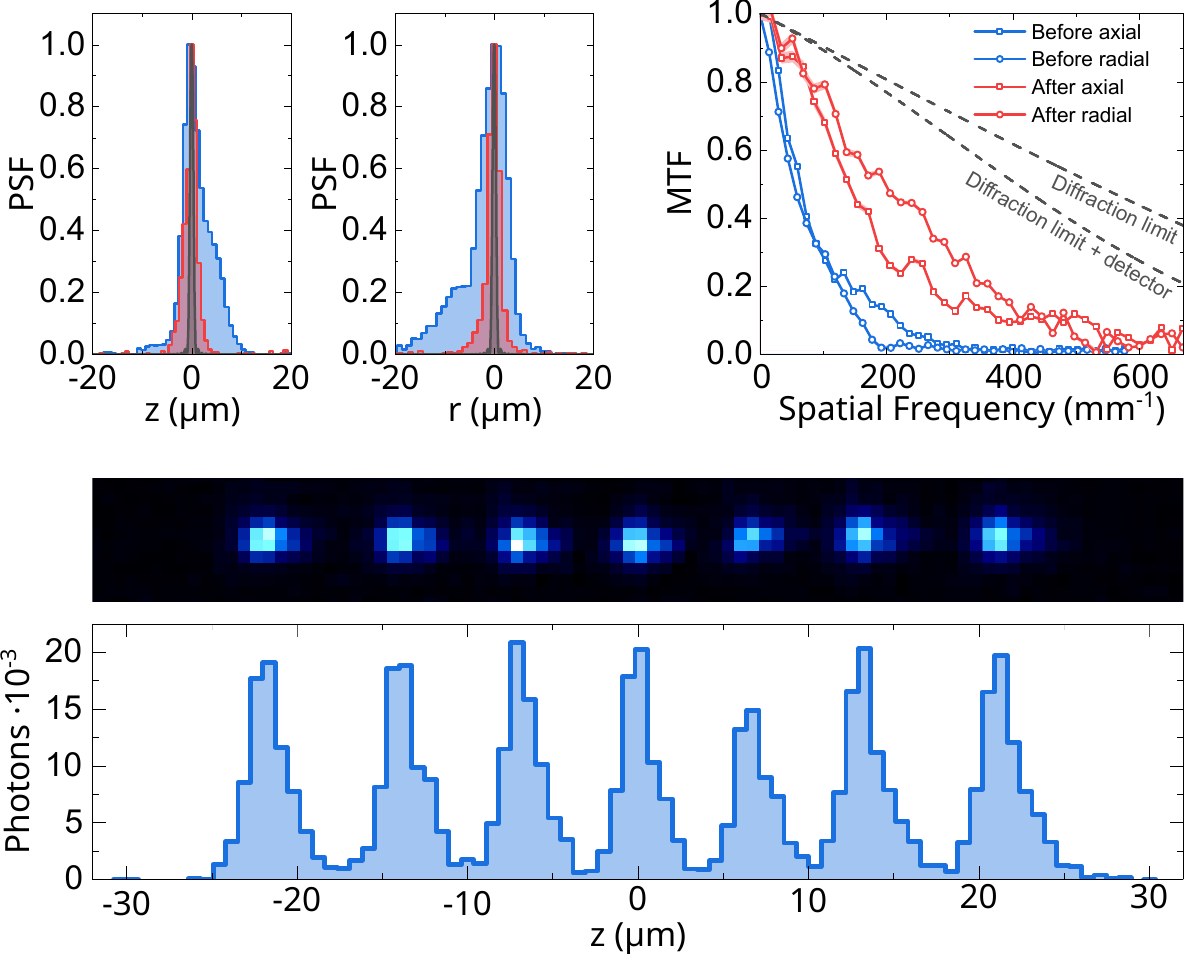}
\caption[Final performance of the optical system]{Final performance of the optical system. Top left:~normalized axial and radial intensity PSF. Top right:~MTF before and after inserting the cylindrical lens compared to the ideal diffraction-limited case. Bottom:~image of a \mbox{seven-ion} linear chain and profile along the axial axis.}
\label{figure_5}
\end{figure}

The intensity PSF analysis of the optical system after the insertion of the cylindrical lens is shown in the lower panels of Fig.~\ref{figure_4}. The first 21 Zernike polynomials were used as in the previous analysis, just to allow a direct comparison. A reduced \mbox{chi-square} of~${\chi_{\nu}=13.1}$ is obtained, although for the fit using the first 15 polynomials this value decreases to ${\chi_{\nu}=11.5}$. The astigmatism is well attenuated compared to the images shown in the upper pannels of Fig.~\ref{figure_4}, and only some of the comma and trefoil components remain significant. The magnification now becomes ${17.9(1)\times}$, larger than before, justified by the translation of the tangential focus point to the sagittal focus point utilizing the cylindrical lens.

Figure~\ref{figure_5} shows the final performance in terms of the PSF and the MTF. Both the images before and after the cylindrical lens inclusion are analyzed. The MTF is calculated directly from the single-ion PSF using FFT~(see Eq.~(\ref{eq:opt_sys_otf}) in Appendix~\ref{appendix_theory}). The maximum spatial frequencies computed, 580~mm$^{-1}$~(before the cylindrical lens correction) and 670~mm$^{-1}$~(after), are limited by the pixel size of 13~$\mu$m. As mentioned in Fig.~\ref{figure_3}, the effect of the detector pixel size is taken into account to determine the real diffraction limit performance~(lower dashed line in Fig.~\ref{figure_5}). The resolution is detailed in Tab.~\ref{tab:opt_sys_resolution}. It should be noted that the intensities are normalized to the maximum value in the~PSF comparison. Finally, a \mbox{seven-ion} linear chain is shown in the lower part of Fig.~\ref{figure_5}. In this particular case, the ions were oscillating at ${\omega_z^\mathrm{com}=2\pi\times333\left(1\right)}$~kHz, resulting in an ion-ion distance in the range \mbox{${d_{\mathrm{ion-ion}}=8.01\left(16\right)}$~-~${6.77\left(17\right)}$~$\mu$m}.

\begin{table}
\caption{Resolution of the optical system before and after inserting the cylindrical lens. These values are calculated from Fig.~\ref{figure_5} using the Rayleigh criterion~(modulation equal to 0.2).}
\begin{tabular}{lrr}
\hline
& \multicolumn{2}{c}{Resolution ($\mu$m)}\\
\cmidrule(lr){2-3}
Situation & Axial & Radial \\\hline
Before & 6.96(5) & 7.89(9) \\
After  & 3.69(3) & 2.75(3) \\
\hline
\end{tabular}
\label{tab:opt_sys_resolution}
\end{table}

\section{Conclusions}

We have designed, characterized, and implemented a new in-vacuum optical system for the detection of single laser-cooled ions using a PMT and an EMCCD in the Penning-trap experiment at the University of Granada. Simulations showing diffraction-limited performance were carried out using the software package Zemax. Out-of-vacuum measurements of the MTF and the CTF using square impulses have shown the near-diffraction-limited performance of the system on the optical axis. For the in-vacuum characterization, a single ion was used as a point-like source to measure the PSF of the system. Wave-aberration retrieval techniques were used to quantify the weight of each type of aberration on the basis of Zernike polynomials. After improving the performance with respect to astigmatism, the resolutions in the axial and radial directions of the trap are 3.69(3)~$\mu$m and 2.75(3)~$\mu$m, respectively.

Motional amplitudes as small as 3~$\mu$m in the radial plane could be detected using the technique described in Ref.~\cite{berrocal2024}. This is advantageous due to the reduction of the systematics associated with the trapping field imperfections in Penning-trap mass spectrometry. A further reduction of these uncertainties can be achieved by accessing the quantum regime, where well-established procedures already used in quantum optics experiments in RF traps could be employed using the PMT of the optical system~\cite{cerrillo2021}. Its high signal-to-noise ratio will be essential to discern between two quantum levels in the timescale of milliseconds.

\bmhead{Acknowledgements}

We acknowledge support from Grant No. PID2019-104093GB-I00 funded by MICIU/AEI /10.13039/501100011033, from FEDER/Junta de Andaluc\'ia - Consejer\'ia de Universidad, Investigaci\'on e Innovaci\'on through Project No. P18-FR-3432, from the Spanish Ministry of Education through PhD fellowship FPU17/02596, and from the University of Granada "Laboratorios Singulares 2020". We thank C. Ospelkaus for providing us access to Zemax, which was utilized for the simulations of the new optical system.  The construction of the facility was supported by the European Research Council (Contract No. 278648-TRAPSENSOR), Projects No. FPA2015-67694-P (funded by MICIU/AEI/10.13039/501100011033 and by ERDF, EU) and No. FPA2012-32076 (MICIU/FEDER), infrastructure Projects No. UNGR10-1E-501, and No. UNGR13-1E-1830 (MICIU/FEDER/UGR), and No. EQC2018-005130-P (funded by MICIU/AEI/10.13039/501100011033 and by ERDF, EU), and infrastructure Projects No. INF-2011-57131 and No. IE2017-5513 (funded by Junta de Andaluc{\'i}a/FEDER).

\begin{appendices}

\section{Technical specification of the lenses}\label{appendix_tables}

The objective lenses' specifications, including one of the field lenses of FL1~(see Fig.~\ref{figure_1}), are detailed in Tab.~\ref{tab:opt_sys_objective}.

\begin{table}[h]
\centering
\caption[Objective's components and relative separation]{Objective's components and relative separation. The distance is defined between adjacent surfaces. All the lenses are commercial models from Thorlabs,~Inc..}
\begin{tabular}{llr}
\hline
Type & Model & Separation (mm)       \\\hline
Object        & -         & 21.963 \\
Asphere       & AL1225G-A & 5      \\
Bi-convex     & LB1844-A  & 6.897  \\
Plano-concave & LC4413-UV & 84.723 \\
Plano-convex  & LA1207-A  & 2.5    \\
Image         & -         & -      \\
\hline
\end{tabular}
\label{tab:opt_sys_objective}
\end{table}

The specifications of the optical elements used for the relays are listed in Tab.~\ref{tab:opt_sys_relay}. These parts are symmetric, so only half of the lenses are detailed. As in the case of the objective, one of the field lenses is included.

\begin{table}[h]
\centering
\caption[Half- and one-inch relay's components and relative separation]{Half- and one-inch relay's components and relative separation. The distance definition between elements is defined in Tab.~\ref{tab:opt_sys_objective}. The lenses are commercial models from Thorlabs,~Inc.~(T) and Newport Corporation~(N).}
\begin{tabular}{lllrr}
\hline
& \multicolumn{2}{c}{Model} & \multicolumn{2}{c}{Separation (mm)} \\
\cmidrule(lr){2-3}\cmidrule(lr){4-5}
Type & 1'' & 1/2''& 1'' & 1/2'' \\\hline
Object        & -             & -            & 2.5    & 3.75    \\
Plano-convex  & LA1207-A (T)  & LA1461-A (T) & 83.496 & 172.666 \\
Plano-concave & LC4413-UV (T) & SPC025 (N)   & 7.227  & 14.814  \\
Bi-convex     & LB1844-A (T)  & LB1676-A (T) & 2.5    & 5       \\
Image         & -             & -            & -      & -       \\
\hline
\end{tabular}
\label{tab:opt_sys_relay}
\end{table}

\section{Evaluation of the performance of an optical system}\label{appendix_theory}

Given the linearity of the electromagnetic field, the image scalar field~${U_i\left(u,v\right)}$ produced by the optical system can be expressed in terms of the object scalar field~${U_{g}\left(\xi,\eta\right)}$ utilizing the superposition integral,
\begin{equation}
\label{eq:sup_int}
U_i\left(u,v\right) = \int\int\limits_{-\infty}^{+\infty} \mathrm{PSF}\left(u,v;\xi,\eta\right) U_g\left(\xi,\eta\right) \,d\xi\,d\eta,
\end{equation}
where~${\mathrm{PSF}\left(u,v;\xi,\eta\right)}$ is the so-called point spread function~(PSF)~\cite{goodman}. If observation takes place on the paraxial image plane, i.e., the lens law is satisfied, the PSF can be expressed in terms of the pupil function of the optical system~$\mathcal{P}\left(x,y\right)$ as
\begin{equation}
\label{eq:opt_sys_psf}
\mathrm{PSF}\left(u,v;\tilde{\xi},\tilde{\eta}\right) = \frac{1}{\lambda^2z_gz_i} \int\int\limits_{-\infty}^{+\infty} \mathcal{P}\left(x,y\right) e^{-i\frac{2\pi}{\lambda z_{i}}\left[\left(u-\tilde{\xi}\right)x+\left(v-\tilde{\eta}\right)y\right]} \,dx\,dy,
\end{equation}
where ${\tilde{\xi}=M\xi,\tilde{\eta}=M\eta}$ being $M$ the magnification, $z_g$ is the longitudinal distance between the object and the entrance pupil, and $z_i$ between the exit pupil and the image plane~\cite{goodman}. Therefore, the PSF is precisely the Fraunhofer diffraction of the generalized pupil function~\cite{goodman}. Equation~(\ref{eq:opt_sys_psf}) also implies the space invariance of the PSF, since ${\mathrm{PSF}\left(u,v;\tilde{\xi},\tilde{\eta}\right)\equiv \mathrm{PSF}\left(u-\tilde{\xi},v-\tilde{\eta}\right)}$. The space-invariant PSF, ${\mathrm{PSF}\left(u,v\right)}$, is the Fourier Transform of the generalized pupil function from the pupil coordinates ${\left(x,y\right)}$~(at the exit pupil) to the spatial frequency coordinates ${\left(\frac{u}{\lambda z_{i}},\frac{v}{\lambda z_{i}}\right)}$~(at the image plane).

In the ideal case of a non-aberrated circularly-symmetric optical system, the wave aberration $W$ is zero, ${\mathcal{P}\left(x,y\right)}$ is constant over the exit pupil, and the PSF is the Airy disc~\cite{bornwolf}. In general, $W$ can be expanded into the Zernike polynomials $Z_{n}^{m}$ as
\begin{equation}
\label{eq:opt_sys_wavefront}
W\left(r,\phi\right)=\sum_{n}\sum_{m=-n}^{n}c_{nm}Z_{n}^{m}\left(r,\phi\right),
\end{equation}
where $r$ and $\phi$ are the radial distance and polar angle in spherical coordinates. The Zernike polynomials constitute a system of orthogonal polynomials in the unit circle that, for low orders, have a direct relation to the primary aberrations:~defocus, astigmatism, comma, spherical, etc. They are defined as~\cite{bornwolf}
\begin{align}
\label{eq:opt_sys_zernike1}
Z_{n}^{m}\left(r,\phi\right) = R_{n}^{m}\left(r\right) \cos(m\phi) \quad\mathrm{for}\quad m<0, \\
Z_{n}^{m}\left(r,\phi\right) = R_{n}^{m}\left(r\right) \sin(m\phi) \quad\mathrm{for}\quad m>0,
\end{align}
where
\begin{equation}
\label{eq:opt_sys_zernike2}
R_{n}^{m}\left(r\right) = \sum_{s=0}^{\frac{n-m}{2}}\left(-1\right)^{s}\frac{\left(n-s\right)!}{s!\left(\frac{n+m}{2}-s\right)!\left(\frac{n-m}{2}-s\right)!} \;r^{n-2s}.
\end{equation}

Coming back to Eq.~(\ref{eq:sup_int}), the image formation of an extended object will greatly depend on whether the source is coherent or incoherent. The intensity in the image plane for the latter can be computed as~\cite{goodman}
\begin{equation}
I_{i}\left(u,v\right) = \int\int\limits_{-\infty}^{+\infty} \left| \mathrm{PSF}\left(u-\tilde{\xi},v-\tilde{\eta}\right) \right|^{2} I_{g}\left(\tilde{\xi},\tilde{\eta}\right) \,d\tilde{\xi}\,d\tilde{\eta}.
\end{equation}
Fluorescence emission by an ion Coulomb crystal is a clear example of incoherent illumination since there is no correlation between the different emission points.

The level of detail that the optical system can reproduce from an object can be quantified by studying the transmission of the spatial frequency~$f$. The modulation transfer function~(MTF) is defined as the absolute value of the normalized Fourier transform of the intensity PSF,
\begin{equation}
\label{eq:opt_sys_otf}
\mathrm{MTF}\left(f_u,f_v\right) = \left|\frac{\int\int\limits_{-\infty}^{+\infty} \left|\mathrm{PSF}\left(u,v\right)\right|^{2} e^{-i2\pi\left(f_uu+f_vv\right)} \,du\,dv}{\int\int\limits_{-\infty}^{+\infty} \left|\mathrm{PSF}\left(u,v\right)\right|^{2} \,du\,dv}\right|,
\end{equation}
and quantifies the relative modulation of the image as a function of the input frequency. It can be demonstrated that aberrations always contribute to decreasing the MTF~\cite{goodman}.

\end{appendices}

\bibliography{bibliography.bib}

\end{document}